\RequirePackage{lineno}
\documentclass[aps,prl,twocolumn,floatfix,showpacs, groupedaddress]{revtex4}

\usepackage{graphicx}
\usepackage{dcolumn}
\usepackage{bm}
\usepackage{multirow}
\usepackage{amsmath}
\usepackage[letterpaper, hmargin = {2cm}, vmargin = {2cm}]{geometry}

\begin{document}


\title{Unexpected distribution of $\nu1f_{7/2}$ strength in the calcium isotopes at $N$=30}

\author{H.~L.~Crawford$^{1,2}$\footnote{Corresponding author: hlcrawford@lbl.gov}}
\author{A.~O.~Macchiavelli$^{1}$}
\author{P.~Fallon$^{1}$}
\author{M.~Albers$^{3}$}
\author{V.~M.~Bader$^{4,5}$}
\author{D.~Bazin$^{4}$}
\author{C.~M.~Campbell$^{1}$}
\author{R.~M.~Clark$^{1}$}
\author{M.~Cromaz$^{1}$}
\author{J.~Dilling$^{6,7}$}
\author{A.~Gade$^{4,5}$}
\author{A.~Gallant$^{6,7}$}
\author{J.~D.~Holt$^{6}$}
\author{R.~V.~F.~Janssens$^{3}$}
\author{R.~Kr\"{u}cken$^{6,7}$}
\author{C.~Langer$^{4}$}
\author{T.~Lauritsen$^{3}$}
\author{I.~Y.~Lee$^{1}$}
\author{J.~Men\'endez$^{8}$}
\author{S.~Noji$^{4}$}
\author{S.~Paschalis$^{9,}$\footnote{Present address: Department of Physics, University of York, York YO10 5DD, United Kingdom}}
\author{F.~Recchia$^{4}$}
\author{J.~Rissanen$^{1}$}
\author{A.~Schwenk$^{9,10}$}
\author{M.~Scott$^{4,5}$}
\author{J.~Simonis$^{9,10}$}
\author{S.~R.~Stroberg$^{4,5,}$\footnote{Present address: TRIUMF, 4004 Wesbrook Mall, Vancouver, British Columbia V6T 2A3, Canada}}
\author{J.~A.~Tostevin$^{11}$}
\author{C.~Walz$^{9}$}
\author{D.~Weisshaar$^{4}$}
\author{A.~Wiens$^{1}$}
\author{K.~Wimmer$^{12}\footnote{Present address: Department of Physics, The University of Tokyo, Hongo, Bunkyo-ku, Tokyo 113-0033, Japan.}$}
\author{S.~Zhu$^{3}$}

\affiliation{$^{1}$Nuclear Science Division, Lawrence Berkeley National Laboratory,
Berkeley, CA 94720, USA}
\affiliation{$^{2}$Institute for Nuclear and Particle Physics and Department of Physics and Astronomy, Ohio University, Athens, OH 45701, USA}
\affiliation{$^{3}$Physics Division, Argonne National Laboratory, Argonne, IL 60439, USA}
\affiliation{$^{4}$National Superconducting Cyclotron Laboratory, Michigan State University,
East Lansing, MI 48824, USA}
\affiliation{$^{5}$Department of Physics and Astronomy, Michigan State University, East Lansing, MI 48824, USA}
\affiliation{$^{6}$TRIUMF, 4004 Wesbrook Mall, Vancouver, British Columbia V6T 2A3, Canada}
\affiliation{$^{7}$Department of Physics and Astronomy, University of British Columbia, Vancouver, British Columbia V6T 1Z1, Canada}
\affiliation{$^{8}$Department of Physics, The University of Tokyo, Hongo, Bunkyo-ku, Tokyo 113-0033, Japan}
\affiliation{$^{9}$Institut f\"ur Kernphysik, Technische Universit\"at Darmstadt, 64289 Darmstadt, Germany}
\affiliation{$^{10}$ExtreMe Matter Institute EMMI, GSI Helmholtzzentrum f\"ur Schwerionenforschung GmbH, 64291 Darmstadt, Germany}
\affiliation{$^{11}$Department of Physics, University of Surrey, Guildford, Surrey GU2 7XH, United Kingdom}
\affiliation{$^{12}$Department of Physics, Central Michigan University, Mt. Pleasant, MI 48859, USA}


\begin{abstract}
The calcium isotopes have emerged as an important testing ground for new microscopically derived shell-model interactions, and a great deal of focus has been directed toward this region.  We investigate the relative spectroscopic strengths associated with $1f_{7/2}$ neutron hole states in $^{47, 49}$Ca following one-neutron knockout reactions from $^{48,50}$Ca.  The observed reduction of strength populating the lowest 7/2$^{-}_{1}$ state in $^{49}$Ca, as compared to $^{47}$Ca, is consistent with the description given by shell-model calculations based on two- and three-nucleon forces in the neutron $pf$ model space, implying a fragmentation of the $l$=3 strength to higher-lying states.  The experimental result is inconsistent with both the GXPF1 interaction routinely used in this region of the nuclear chart and with microscopic calculations in an extended model space including the $\nu1g_{9/2}$ orbital.
\end{abstract}

\maketitle


The calcium isotopic chain is a present focus of nuclear structure physics, both from experimental and theoretical perspectives.  This isotopic chain contains novel and intriguing examples of evolving shell structure far from stability~\cite{Wienholtz2013,Steppenbeck2013} and is an active region for the study and test of three-body (3N) forces used in microscopically derived shell-model interactions and large-space ab initio calculations~\cite{Wienholtz2013,Holt2012,Hagen2012,Gallant2012,Holt2013,Hagen2013,Soma2014,Binder2014,Holt2014,Hergert2014,Hebeler2015,Hagen2015,Garcia2016,Hagen2016,Stroberg2016}.

From the theoretical perspective, new developments are enabling a microscopic description of these nuclei, with calculations being performed from $^{48}$Ca to $^{70}$Ca using effective shell-model interactions~\cite{Holt2012, Holt2014}, or large-space calculations~\cite{Hagen2012, Soma2014, Binder2014, Hergert2014, Hagen2015, Stroberg2016} based on two-nucleon (NN) and 3N interactions derived from chiral effective field theory.  These calculations have already shown differences compared to predictions of phenomenologically based shell-model interactions, even for nuclei as close to stability as $^{50}$Ca.  For larger neutron number $N$, predictions for the location of the dripline are strongly model dependent, varying from $^{60}$Ca to $^{76}$Ca~\cite{Holt2012, Hagen2012, Erl2012}.  Data on the structure of the neutron-rich Ca isotopes is critical to benchmark the various new calculations and validate their predictions.

Measurements of properties such as masses~\cite{Gallant2012, Wienholtz2013} and spectroscopy~\cite{Steppenbeck2013} at the limits of current facilities are well reproduced by the newest calculations, but recent data have revealed discrepancies with theoretical predictions, bringing into question extrapolations toward the dripline.  For example, a laser spectroscopy measurement at CERN-ISOLDE~\cite{Garcia2016} reported charge radii that show an anomalously large increase from $^{48}$Ca to $^{52}$Ca, which significantly exceeds all theoretical predictions.
 
Single-particle occupancies, while not observables, can provide a 
test for theoretical descriptions.
Phenomenological interactions like GXPF1 \cite{Honma2002,Honma2005} and microscopically based interactions both find reasonable agreement with spectroscopic data, but they predict different distributions of the neutron $\nu1f_{7/2}$ strength in $^{49}$Ca.  Phenomenological models are more consistent with the single-particle description, where one would expect the full $\nu1f_{7/2}$ strength to be concentrated in the lowest 7/2$^{-}_{1}$ state for Ca nuclei at and immediately beyond $N$=28.  However, the microscopic interactions suggest a possible fragmentation of this $\nu1f_{7/2}$ strength.  

In this Letter, we report the results of a neutron-knockout (-1n) experiment performed in the Ca isotopes at $N$=28 and $N$=30, using the high-resolution $\gamma$-ray detection array GRETINA~\cite{GRETINA} to measure exclusive neutron-knockout cross sections from $^{50}$Ca to states in $^{49}$Ca, and from $^{48}$Ca to $^{47}$Ca.  Based on the data and theoretical cross sections, calculated under the assumption of the sudden removal of a neutron with a given set of quantum numbers~\cite{Tostevin2014, Gade2005}, we extract spectroscopic factors, whose sum rule gives the occupancy of a given neutron single-particle orbital.  A relative measurement, such as that performed here comparing $^{48}$Ca(-1n) and $^{50}$Ca(-1n) neutron removal, provides a framework to firmly establish the trend in the spectroscopic strength distributions for the neutron $pf$ orbitals.  Our results indicate a decrease in the population of the lowest 7/2$^{-}$ state in $^{50}$Ca,
at odds with the phenomenological description. This trend is partially reproduced by NN+3N calculations in the $pf$ shell-model space.


The experiment was performed at the National Superconducting Cyclotron Laboratory (NSCL) at Michigan State University.  Secondary beams of $^{48,50}$Ca were produced following fragmentation of a 140-MeV/$u$ $^{82}$Se primary beam on a 423~mg/cm$^{2}$ $^{9}$Be target.  Reaction products were then separated through the A1900 fragment separator~\cite{A1900}, based on magnetic rigidity and relative energy loss through a 600~mg/cm$^{2}$ Al degrader wedge.  Fragments were delivered with a momentum acceptance of 2\% $\Delta$p/p and impinged on a 370~mg/cm$^{2}$-thick Be reaction target located at the target position of the S800 spectrograph~\cite{S800}. The resulting knockout products were identified on an event-by-event basis through time-of-flight and energy loss as measured by the focal plane detectors of the S800.

\begin{figure*}[htbp]
\includegraphics[width=\textwidth]{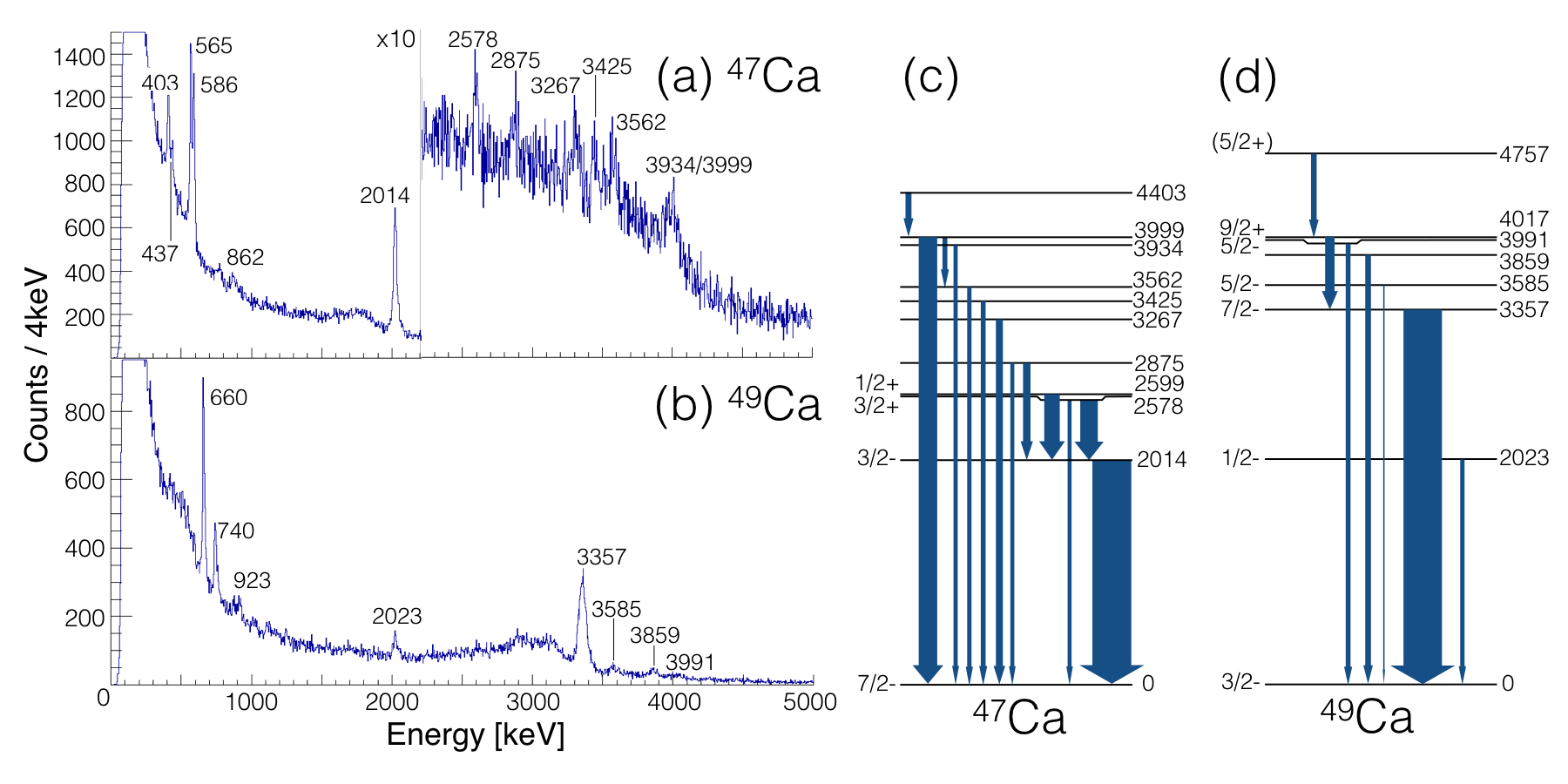}
\caption{Experimentally observed $\gamma$-ray spectra (left) and populated level schemes (right) of (a) $^{47}$Ca from $^{48}$Ca(-1n), and (b) $^{49}$Ca from $^{50}$Ca(-1n) reactions.  Experimental spectra are marked with the observed transitions in units of keV.  In the level schemes presented in (c) and (d) the assigned spins and parities come from the literature, while the width of the arrows corresponds to the (efficiency-corrected) relative intensity of the $\gamma$-ray transitions.  For reference, $S_{n}$($^{47}$Ca) = 7.3~MeV and $S_{n}$($^{49}$Ca) = 5.1~MeV.}
\label{fig:spectraAndLevels}
\end{figure*}

Seven GRETINA~\cite{GRETINA} modules surrounded the target position of the S800 and were used to detect $\gamma$ rays emitted from excited states populated in the knockout residues.  Four modules were placed at forward ($\theta\sim$ 58$^{\circ}$) angles, and three at $\theta\sim$ 90$^{\circ}$ relative to the beam direction.  Each GRETINA module consists of four closely packed, high-purity germanium crystals (28 in total), with each crystal electronically segmented into 36 individual elements.  The degree of segmentation combined with the decomposition of the full set of signals allows the positions and energies of individual $\gamma$-ray interaction points to be measured.  The $\gamma$-ray interaction position information from GRETINA, along with the particle trajectory information from the S800 were used to provide an accurate event-by-event Doppler reconstruction of the observed $\gamma$ rays.  An overall $\gamma$-ray resolution of $\sim$1.5\% was achieved following Doppler correction.  Yields for individual transitions were determined by a fit to data using a GEANT4 simulation of the GRETINA response~\cite{Riley2016}, including the angular distribution of emitted $\gamma$ rays (based on the calculated population of the $m$ substates in the knockout reaction); the simulation is conservatively taken to contribute an absolute error of 1\% to the $\gamma$-ray efficiency.  The results are summarized in Fig.~\ref{fig:spectraAndLevels} and Table~\ref{tab:crossSections}. 


The Doppler-shift corrected spectra of $\gamma$-rays detected in GRETINA in coincidence with $^{47,49}$Ca reaction products
are presented in Figs.~\ref{fig:spectraAndLevels}(a) and (b).  The corresponding level schemes (excited states and transitions), as observed in this work, are
shown in Figs.~\ref{fig:spectraAndLevels}(c) and (d). Thirteen transitions are observed and associated with levels in $^{47}$Ca populated in the one neutron removal reaction. The majority of the transitions were previously observed~\cite{Burrows2007} and their placement in the level scheme follows the literature.  Two $\gamma$ rays, at 3425 and 3267~keV were not previously reported, but are placed as transitions directly to the ground state, supported by $\beta$-decay data~\cite{Smith2016}.  Where statistics are sufficient, the level scheme was verified by $\gamma$-$\gamma$ coincidences.
For $^{48}$Ca(-1n), the states of primary interest are at 2.59, 2.58 and 0~MeV (ground state) corresponding to direct removal of a $s_{1/2}$, $d_{3/2}$ and $f_{7/2}$ neutron,  respectively. 

For the case of $^{49}$Ca, eight transitions of appreciable statistics are observed, all of which have been previously placed in the level scheme~\cite{Montanari2011, Broda2001}.
Here, the 3.4~MeV state and the ground state are of primary interest, corresponding to the removal of a $f_{7/2}$ and $p_{3/2}$ neutron, respectively.  We note that the 1/2$^{-}$ state at 2.0~MeV may also be populated through direct removal of a $p_{1/2}$ neutron, should such a configuration be present in the $^{50}$Ca ground state.

Cross sections for the direct population of the states of interest in $^{47,49}$Ca are given in Table~\ref{tab:crossSections} and 
were deduced from the observed level schemes while accounting for feeding from higher-lying states.  Cross sections were also corrected for losses associated with the momentum acceptance of the S800.  The exclusive parallel momentum distributions were found to be consistent with the expected angular momentum transfer for these states;  i.e., $l$ = 0, 2, and 3 for the 2.59, 2.58~MeV, and ground state, respectively, in  the $^{48}$Ca(-1n) reaction and  $l$ = 3 and 1 for the 3.4~MeV and ground state in the $^{50}$Ca(-1n) reaction.  A fit of the partial momentum distributions with the calculated distributions allowed us to deduce the required acceptance correction factors, on average contributing a correction of order 10\%, with a maximum value of 20\%.  These corrections contributed an error of 10\% to the overall error budget.

The measured neutron knockout cross sections can be compared to the various theoretical predictions and used to assess them.
They can also be used to derive an occupancy; i.e., the number of neutrons in a particular single-particle state.
To compare to theory and relate the measured cross section to an occupancy, we first calculate a theoretical single-particle knockout cross section,  $\sigma_{sp}$, corresponding to the removal of a neutron with a given set of quantum numbers and assuming a spectroscopic factor, C$^{2}$S, of 1.  Details of the formalism and methodology used to calculate $\sigma_{sp}$ are given in Refs.~\cite{Tostevin2014, Gade2005, Gade2008}.  From the calculated $\sigma_{sp}$ and measured cross sections,  $\sigma_{-1n}$, given in Table~\ref{tab:crossSections}, we extract experimental spectroscopic factors C$^{2}$S$_{exp}$.

\begin{table*}[!hbtp]
\caption{Partial summary of states populated in the one-neutron removal reactions from $^{48}$Ca$\rightarrow^{47}$Ca and $^{50}$Ca$\rightarrow^{49}$Ca.  Level energies and spin/parity assignments are taken from the literature~\cite{Broda2001, Burrows2007, Montanari2011}.  Single-particle theoretical cross-sections,  $\sigma_{sp}$, along with an $A$-dependent center-of-mass correction~\cite{Dieperink1974} are used to deduce the values for C$^{2}$S$_{exp}$ shown.  $R_{S}$ quenching factors are extracted based on a fit to systematics~\cite{Tostevin2014} and are used to calculate C$^{2}$S$^{norm}_{exp}$ (see text for details).  Theoretical values are provided for the phenomenological GXPF1 shell-model interaction, as well as the NN+3N-based interaction in the $pf$ and $pfg_{9/2}$ model spaces.
The range of values for the NN+3N cases is an estimate for the uncertainty associated with varying the NN cutoff in the derivation of the interaction.}
\label{tab:crossSections}
\renewcommand{\arraystretch}{1.2}
\begin{tabular}{cccccc|c|ccc}
\hline\hline
Level Energy & \multirow{2}{*}{J$^{\pi}$} & $\sigma_{-1n}$ & 
$\sigma_{sp}$ & \multirow{2}{*}{C$^{2}$S$_{exp}$} & \multirow{2}{*}{R$_{S}$} & \multirow{2}{*}{C$^{2}$S$_{exp}^{norm}$} & \multicolumn{3}{c}{Theoretical C$^{2}$S }\\
$[$keV] &  & [mb] & 
[mb] & & 
 & & GXPF1 & \textit{pf} NN+3N & \textit{pfg$_{9/2}$} NN+3N\\
\hline
\multicolumn{10}{c}{$^{48}$Ca$\rightarrow^{47}$Ca}\\
\hline
0 & 7/2$^{-}$ & 70.6$^{+8.4}_{-9.6}$ & 11.01 & 6.4$^{+0.8}_{-0.9}$ & 0.69 & 9.3$(^{+1.2}_{-1.3})_{stat}(\pm$1.9$)_{sys}$ & 7.7 & $7.2-7.4$ & $6.7-7.0$\\
2014 & 3/2$^{-}$ & $\leq$1.4 & 11.24 & $\leq$0.2 & 0.66 & $\leq$0.2 & 0.06 & $0.05-0.07$ & $0.05-0.07$ \\
2578 & 3/2$^{+}$ & 9.4$^{+3.1}_{-1.9}$ & 7.46 & 1.2$^{+0.4}_{-0.2}$ &  0.65 & 1.8($^{+0.6}_{-0.3})_{stat}(\pm0.4)_{sys}$& \\
2599 & 1/2$^{+}$ & 10.5$^{+1.4}_{-1.3}$ & 12.58 & 0.8(1) & 0.65 & 1.2$(\pm0.2)_{stat}(\pm0.2)_{sys}$ & \\
\hline
Inclusive & & 123(10)\\
\hline
\multicolumn{10}{c}{$^{50}$Ca$\rightarrow^{49}$Ca}\\
\hline
0 & 3/2$^{-}$ & 41.8$^{+5.2}_{-5.9}$ & 18.63 & 2.1(3) & 0.77 & 2.7$(\pm0.4)_{stat}(\pm0.5)_{sys}$ & 1.73 & $1.70-1.72$ & $1.50-1.56$ \\
2023 & 1/2$^{-}$ & 4.4$^{+0.8}_{-0.5}$ & 15.04 & 0.3(1) & 0.74 & 0.4$(\pm0.1)_{stat}(\pm0.1)_{sys}$ & 0.17 & $0.12-0.14$ & $0.12-0.14$ \\
3357 & 7/2$^{-}$ & 38.9$^{+5.1}_{-3.9}$ & 10.87 & 3.4$^{+0.4}_{-0.3}$ & 0.72 & 4.7$(^{+0.6}_{-0.4})_{stat}(\pm0.9)_{sys}$ & 7.7 & $5.6-5.7$ & $6.3-6.7$ \\
4017 & 9/2$^+$ & 4.1(8) & -- & -- & -- & -- & -- & -- & $0.15-0.20$\\
\hline
Inclusive & & 116(8)\\
\hline
\hline
\end{tabular}
\end{table*}

To compare with the results of shell-model calculations, we take the further step of correcting by $R_{S}$, a suppression or quenching factor, known to be required to scale calculated single-particle cross sections to measurements \cite{Tostevin2014, Gade2008}.
Following the analysis of Ref.~\cite{Mutschler2016}, the $R_{S}$ values used here and given in Table~\ref{tab:crossSections} were obtained from a fit to the systematics of $R_{S}$ as a function of $\Delta S = S_{n} - S_{p}$ for inclusive neutron knockout data~\cite{Tostevin2014}.  We assume a 20\% systematic error associated with the local scatter in $R_{S}$ as a function of $\Delta S$, which is propagated in the calculation of C$^{2}$S$_{exp}^{norm}$.

The value of C$^{2}$S$_{exp}^{norm}$ = 9.3$(^{+1.2}_{-1.3})_{stat}(\pm1.9)_{sys}$ for the lowest $1f_{7/2}$ state in $^{47}$Ca is consistent with the results obtained in (p, d) and (d, t) neutron transfer measurements~\cite{Martin1972, Williams1977} and with the expected value of 8 (i.e., a full $1f_{7/2}$ orbital in the $^{48}$Ca ground state).  The spectroscopic factors to the lowest $1d_{3/2}$ state at 2.58~MeV and the lowest $2s_{1/2}$ state at 2.60~MeV in $^{48}$Ca(-1n) are similarly consistent with the literature values.
However, in $^{50}$Ca(-1n) the spectroscopic factor to the first 7/2$^{-}$ state at 3.36~MeV is significantly lower than that observed in $^{48}$Ca(-1n), at only 4.7$(^{+0.6}_{-0.4})_{stat}(\pm0.9)_{sys}$.  The C$^{2}$S$_{exp}^{norm}$ values for the population of the ground state ($\nu2p_{3/2}^{1}$ level) in $^{49}$Ca, and the first excited state at 2.0~MeV ($\nu2p_{1/2}^{1}$ state) are 2.7$(\pm0.4)_{stat}(\pm0.5)_{sys}$ and 0.4$(\pm0.1)_{stat}(\pm0.1)_{sys}$, respectively.    

In this work we update the calculations of Refs.~\cite{Holt2012,Holt2013,Holt2014} to compare with these new experimental measurements. Following the same perturbative many-body approach for generating the $pf$ and $pfg_{9/2}$ valence-space Hamiltonians outlined in Ref.~\cite{Holt2014}, we start from NN+3N interactions that predict realistic saturation properties of nuclear matter within theoretical uncertainties \cite{Hebeler2011,Simonis2016}. These interactions have also been recently used to study the Ca isotopes \cite{Hagen2015,Garcia2016}. By varying the low-resolution cutoff in NN forces from $\lambda_{\mathrm{NN}}\!=\!1.8\!-\!2.2$ fm$^{-1}$, we obtain an uncertainty estimate for the calculations.  Low-lying excited states for the $pf$-shell calculation agree reasonably well with experiment.
For instance in $^{47}$Ca, the $3/2^-_1$ and $7/2^-_1$ states lie at 2.15 MeV and 3.23 MeV excitation energy, respectively, for $\lambda_{\mathrm{NN}}=1.8$ fm$^{-1}$,
within 200 keV of experiment, and $1/2^-_1$ and $3/2^-_2$ states are predicted
below 3 MeV, in agreement with spin-unassigned experimental levels.
All states are shifted approximately 300 keV and 600 keV higher in energy for $\lambda_{\mathrm{NN}}=2.0,2.2$ fm$^{-1}$.
In $^{49}$Ca the central energy values given by $\lambda_{\mathrm{NN}}=2.0$ fm$^{-1}$ of $E(1/2^-_1)=2.07(05)$ MeV, $E(5/2^-_1)=2.32(03)$ MeV, $E(7/2^-_1)=3.40(30)$ MeV, and $E(5/2^-_1)=3.53(25)$ MeV, with approximate uncertainties in parentheses, agree well with experiment outside of the $5/2^-_1$ state, predicted more than
1 MeV too low in energy.  In all cases excited states calculated in the $pfg_{9/2}$ valence-space lie within 200 keV of their $pf$-shell counterparts,
suggesting that the $g_{9/2}$  orbital does not play a big role for these isotopes.

\begin{figure}[htbp]
\includegraphics[width=0.5\textwidth]{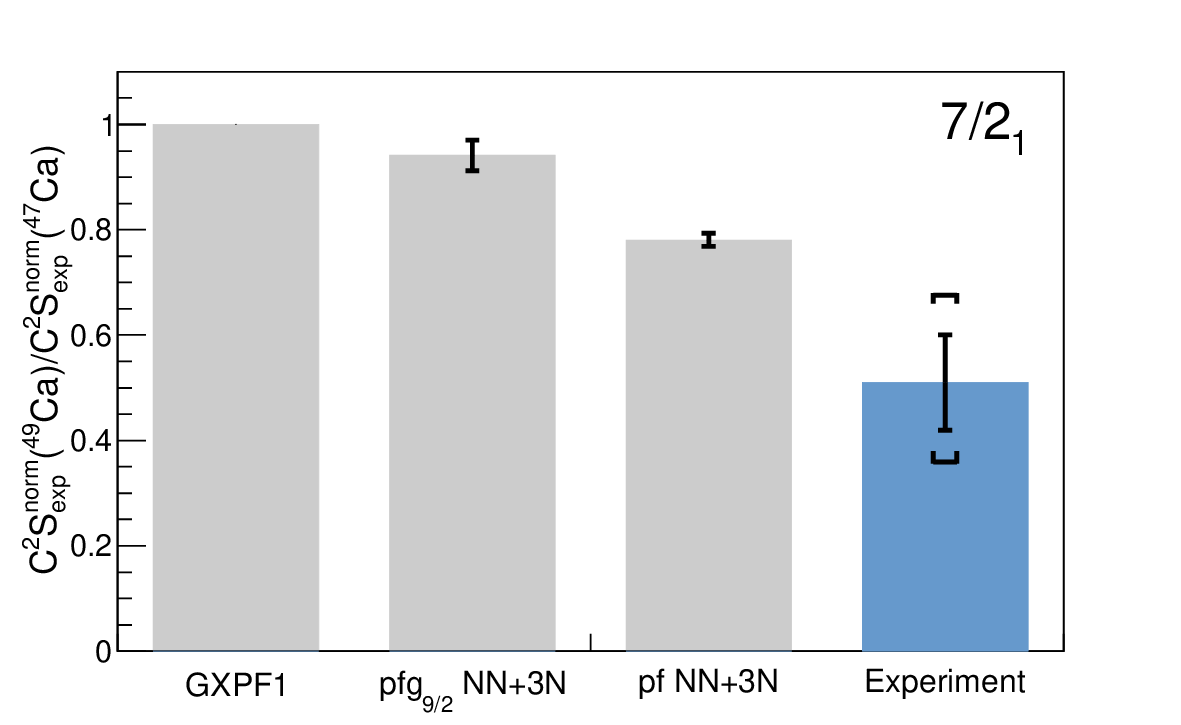}
\caption{Ratio of the spectroscopic factor for the population of the first 7/2$^{-}$ state in one-neutron knockout from $^{50}$Ca$\rightarrow^{49}$Ca / $^{48}$Ca$\rightarrow^{47}$Ca, as predicted by the GXPF1A phenomenological shell-model interaction (left), shell-model calculations based on NN+3N forces in the \textit{pf} (center right) and \textit{pfg$_{9/2}$} (center left) model spaces, and measured in the present experiment (blue column, right).}
\label{fig:results}
\end{figure}

The ratio of spectroscopic factors to populate the lowest 7/2$^{-}_{1}$ state in the neutron knockout from $^{50}$Ca and $^{48}$Ca is plotted in Fig.~\ref{fig:results}. 
It is evident that there is marked difference between the
experimental and theoretical ratios,
which, however, become almost compatible when considering respective uncertainties.
In the phenomenological description, the full $1f_{7/2}$ strength of C$^{2}$S = 8 is concentrated in the lowest 7/2$^{-}$ state in both the $^{48}$Ca and $^{50}$Ca reactions.  For the NN+3N calculations, the $1f_{7/2}$ strength is also largely concentrated in the 7/2$^{-}_{1}$ state at $N$=28, but in $^{50}$Ca(-1n), a reduced strength to the 7/2$^{-}_{1}$ state is seen, particularly for the $pf$ valence-space calculation.
Consequently, both GXPF1 and the \textit{pfg$_{9/2}$}NN+3N interaction predict a ratio $\approx$ 1, while for the $pf$ interaction the ratio is 0.78.  Experimentally, we determine a ratio of 0.51$(\pm0.09)_{stat}(\pm0.15)_{sys}$, shown by the blue bar in Fig.~\ref{fig:results}. The error bar indicates the statistical error from the data; the bracketed error bar represents the systematic error associated with the determination of $R_{S}$.  

The disagreement with the well-established phenomenological GXPF1 interaction can provide important feedback to refine this family of interactions.  
Likewise, disagreement with the \textit{pfg$_{9/2}$} NN+3N predictions
along with deficiencies in spectroscopy of low-lying 9/2$^{+}$ states \cite{Montanari2011, Gade2016}
may call for an improved treatment in valence spaces beyond one major shell \cite{Tsunoda2014}.
The most reasonable agreement is found for the the $pf$ NN+3N interaction. 
The reduced cross section in the $^{50}$Ca(-1n) reaction is due to a fragmentation of the $1f_{7/2}$ strength to states at higher excitation energies in $^{49}$Ca, as of yet unidentified and potentially above the neutron separation energy.
Nevertheless, it is worth noting that the extent of the reduction is larger in experiment than the predictions of the $pf$-shell NN+3N calculation.


Finally, we comment briefly on the $^{50}$Ca(-1n) spectroscopic factor populating the $^{49}$Ca ground state,  associated with removal of $2p_{3/2}$ neutrons.  Within an extreme single-particle description, a value of C$^{2}$S = 2 is expected --  both GXPF1 and the two NN+3N interactions exhaust $>$75\% of this maximum value.  Including the possible systematic error associated with overestimation of the ground state, as discussed above, the present measurement of 2.7$(\pm0.4)_{stat}(\pm0.5)_{sys}$ is slightly above 2, but agrees within errors.  It is interesting to note, however, an apparent enhancement of $l=1$ and depletion of the $l=3$ strength was reported in the neighboring Sc isotopes~\cite{Schwertel2012}.
However, it is clear that no firm conclusions regarding a possible enhancement of $2p_{3/2}$ neutrons in the $^{50}$Ca ground state can be drawn from the present data, leaving an open question regarding the occupancy of the $2p_{3/2}$ neutron orbital.


In summary, current state-of-the-art nuclear shell-model calculations make significantly different predictions regarding the population of
7/2$^{-}_{1}$ states following direct neutron removal from $^{48}$Ca and $^{50}$Ca.
The present results of the $^{48,50}$Ca(-1n) one-neutron knockout reactions performed at NSCL using the high resolution gamma-ray tracking GRETINA, indicate a reduction of the strength populating the lowest 7/2$^{-}_{1}$ state in $^{49}$Ca as compared to $^{47}$Ca.
The data are in best agreement with shell-model calculations based on NN+3N forces in the neutron $pf$ model space, and
are not consistent with calculations using the phenomenological GXPF1 interaction, nor NN+3N calculations including the $\nu1g_{9/2}$ orbital. 
In this context, it is interesting to note that the $\nu1g_{9/2}$ orbital plays an important role in calculations when predicting the location of the neutron dripline in calcium isotopes. 

\begin{acknowledgments}
The authors thank the operations team at NSCL for their work in beam
delivery during the experiment.  This material is based upon work
supported by the U.S. Department of Energy, Office of Science, Office
of Nuclear Physics under Contracts No. DE-AC02-05CH11231 (LBNL) and
No. DE-AC02-06CH11357 (ANL), by the Department of Energy National
Nuclear Security Administration under award number DE-NA0000979, the
National Science Foundation (NSF) under PHY1102511, the European
Research Council Grant No.~307986 STRONGINT and the BMBF under
Contract No.~05P15RDFN1.  HLC also acknowledges the U.S. DOE under
grant No. DE-FG02-88ER40387 (OH).  JAT acknowledges the support of the
United Kingdom Science and Technology Facilities Council (STFC) under
Grant No. ST/L005743/1. JM was supported by an International Research
Fellowship from JSPS, and Grant-in-Aid for Scientific Research
No.\ 26$\cdot$04323.  GRETINA was funded by the U.S. DOE Office of
Science.  Operation of the array at NSCL is supported by NSF under
Cooperative Agreement PHY11-02511 (NSCL) and DOE under Grant
No. DE-AC02-05CH11231 (LBNL). Computations were performed with an
allocation of computing resources at the J\"ulich Supercomputing
Center.
\end{acknowledgments}

\end{document}